\documentclass[proceedings, preprint]{rmaa}

\usepackage{paralist}

\usepackage{psfrag,color}

\newcommand{\pasa}{Publications of the Astronomical Society of Australia}
  
\SetYear{2024}
\SetConfTitle{Astrorob 2023}

\title{Management of a Multi-User Robotic Observatory} 

\author{
  John Moore,\altaffilmark{1,2} 
	Bruce Gendre,\altaffilmark{1,3} 
	David Coward,\altaffilmark{1} 
	Fiona Panther,\altaffilmark{1,4}
	and Eloise Moore\altaffilmark{1}}

\altaffiltext{1}{University of Western Australia, OzGrav ARC Centre of Excellence, 35 Stirling Highway, M013, 6009 Crawley, WA, Australia.}
\altaffiltext{2}{Email address: john.moore@uwa.edu.au}
\altaffiltext{3}{Present address: University of the Virgin Islands, VI, USA}
\altaffiltext{4}{Forest Fellow.}

\shortauthor{Moore, et al.}
\shorttitle{Zadko Observatory Management}

\listofauthors{J. Moore, B. Gendre, D. Coward, F. Panther, E. Moore}
\indexauthor{Gendre, B.}
\indexauthor{Coward, D.}
\indexauthor{Moore, E.}
\indexauthor{Moore, J.}
\indexauthor{Panther, F.}

\abstract{The Zadko Observatory located approximately 70 kilometres north of Perth in the Yeal nature reserve within the Shire of Gingin, Western Australia, initially housed the 1.0 metre f/4 fast-slew Zadko Telescope which was commissioned in June 2008. Since the Zadko telescope has been in operation it has proven its worth by detecting numerous Gamma Ray Burst afterglows, two of these being the most distant ‘optical transients’ imaged by an Australian telescope. The Zadko telescope also contributed to the discovery of colliding neutron stars in 2017 capturing the imagination of the public. Another important use for the Zadko Telescope is the tracking and mapping of Space Debris which consist of all man-made objects, including their fragments or parts, other than active space vehicles larger than 10 microns and orbiting the Earth in outer space. The Zadko telescope forms part of the ARC Centre of Excellence for Gravitational Wave Discovery (OzGrav). With a view to supporting ongoing scientific instrument upgrades and observatory maintenance it has proven critical to attract additional funding and seek out collaboration with international partners. In this article, we will focus on the administrative and technical details of the Observatory, focusing on the sustainability of the Observatory and its ongoing potential for future growth into an internationally recognized Space Surveillance Hub. We will review the evolution of the Observatory, from its early, single instrument state, to its current multi-telescope and multi-instrument capabilities. We will finish by outlining the future of the Observatory and surrounds.}

\resumen{El Observatorio Zadko, ubicado aproximadamente a 70 km al norte de Perth, en la reserva natural de Yeal dentro de la Comarca de Gingin, Australia Occidental, alberg\'o inicialmente el Telescopio Zadko de rápido movimiento y con un di\'ametro de 1,0 metros a f/4, que se puso en servicio en junio de 2008. Desde entonces, el telescopio Zadko ha demostrado su eficacia al detectar numerosos estallidos de explosiones de rayos gamma, dos de los cuales han sido las \&quot;fuentes transitorias \'opticas\&quot; más distantes fotografiadas por un telescopio australiano. El telescopio Zadko también contribuy\'o al descubrimiento de la fusi\'on de dos estrellas de neutrones en 2017, hecho que tuvo gran repercusi\'on. Otro uso importante del Telescopio Zadko es el seguimiento y cartografiado de desechos espaciales, que consisten en todos los objetos fabricados por el hombre, incluidos sus fragmentos o partes, distintos de los vehículos espaciales activos y con un tama\~no de más de 10 micras y que orbitan la Tierra en el espacio exterior. El telescopio Zadko forma parte del Centro de Excelencia ARC para el Descubrimiento de Ondas Gravitacionales (OzGrav). Con miras a apoyar las actualizaciones en curso de los instrumentos científicos y el mantenimiento de los observatorios, ha resultado fundamental atraer financiación adicional y buscar colaboración con socios internacionales. En este artículo nos enfocamos en los detalles administrativos y técnicos del Observatorio, centr\'andonos en la sostenibilidad del Observatorio y su potencial continuo para el crecimiento futuro hasta convertirse en un Centro de Vigilancia Espacial reconocido internacionalmente. Tambi\'en se comenta la evolución del Observatorio, desde su estado inicial albergando un \'unico instrumento, hasta sus capacidades actuales con m\'ultiples telescopios e instrumentos, para terminar esbozando el futuro del Observatorio y sus alrededores.}

\addkeyword{Miscellaneous}
\addkeyword{Telescopes}

\begin{document}
\maketitle

\section{Introduction}
\label{sec:intro}

\begin{figure*}[!hbt]
  \includegraphics[width=\textwidth]{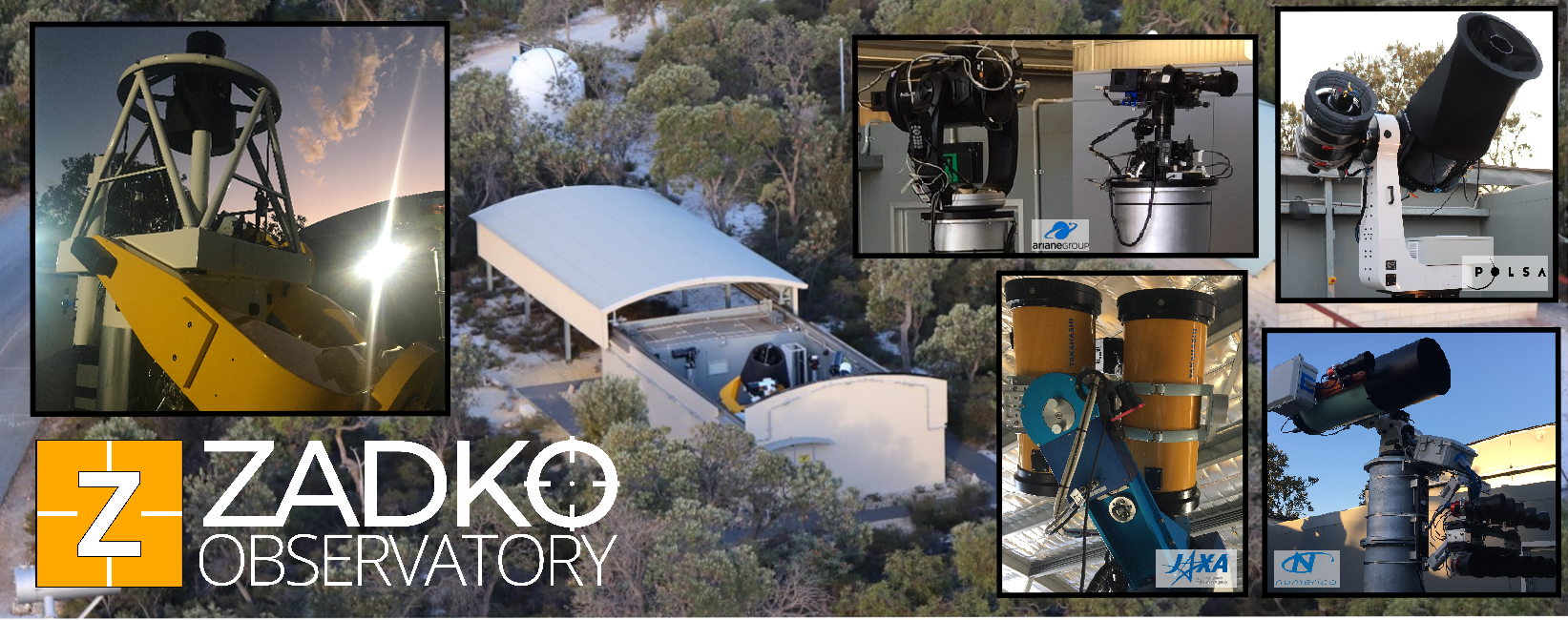}
  \caption{An overview of the Zadko Observatory showing an aerial shot with the rolling roof open and various images of instruments within including (from left to right) the Zadko, Ariane Group, POLSA-1, JAXA and Slingshot telescopes, hosted within the Observatory.}
  \label{fig:one}
\end{figure*}

The Zadko Observatory is a unique fully automated robotic observatory \citep{moo21} located at the Gingin Gravity Precinct, Western Australia (see Figure \ref{fig:one}). The Observatory is of National and International importance, with more than 100 scientists utilizing it routinely and it remains the only facility operating a metre class telescope with deep imaging capability between the east coast of Australia and South Africa in the Southern hemisphere. Indeed, this facility is the only observatory worldwide able to study Southern astronomical events after 02:00 AWST, placing it in high demand both Nationally and Internationally \citep{cow17}. 

Due to the Zadko Observatory’s longitude, it has become apparent that many Eastern States’ Universities and International Space Agencies are keen to utilize our unique geographical location. This is due to the West coast being the last place in Australia to observe the night sky in the early morning: it is still night in Perth while it is already the day on the East coast. For this reason alone, the Zadko Observatory has been identified as a premium site by other astronomical groups and commercial entities to compliment sites located in the East. As a consequence, this can only lead to further demand and development of the Space Surveillance Hub at Gingin. 

The purpose of this short paper is to highlight the management of a multi-user robotic observatory, and how the various conflicting scientific goals are balanced in order to extract as much data from the site as possible. In a companion paper (Gendre et al., this proceeding), we will present how the site location can impact the scientific goals and how we try to mitigate some of them. This paper is organised as follow. In Section \ref{sec:histo} we recall the historical aspects of the observatory, from its initial design to its current one. In Section \ref{sec:leif} we present the current upgrade of the observatory and how we will use the Zadko Telescope. Finally, in Section \ref{sec:future}, we indicate how we expect the Observatory to grow.

\section{Historical aspects}
\label{sec:histo}

The 1.0 metre f/4 Ritchey-Chrétien Zadko Telescope was constructed by DFM Engineering Inc. in Longmont Colorado USA and donated to the University of Western Australia by Resource Company Claire Energy CEO, Jim Zadko. The Zadko Observatory is located at Longitude 115\arcdeg 42\arcmin 49\arcsec E and Latitude 31\arcdeg 21\arcmin 24\arcsec S with Observatory Code D20.
Due to operational issues, the original 6.7 metre fibreglass domed observatory, which the Zadko Telescope was initially housed, was replaced in 2011 with a state of the art purpose built robotic rolling roof autonomous observatory with a dedicated 21 m$^2$ constant temperature climate controlled operations room. In addition to this, a climate controlled telescope service room to mirror the operation room was also added. This had the benefit of allowing for possible future conversion into a second control room when other telescopes were added to the increased 63 m$^2$ telescope observation room. Robotic control of the rolling roof is performed using a PLC-Burgess system designed and manufactured by the electronics workshop at Observatoire de Haute-Provence, France. A brief system overview is explained here: The PLC is connected to a dedicated server via an Ethernet connection. The weather station and the cloud sensor are connected to the serial ports of the server. The PC processes this data and sends commands to the PLC in order to control the roof in a safe condition. The PLC receives these commands from the server and sends its status through a socket connection. All the safety devices (rain detectors, humidity and temperature sensors, power probe before the UPS unit) are controlled by the PLC. In the event of a communication failure with the PC, the PLC closes the roof. 

\section{Current upgrade of the observatory}
\label{sec:leif}

To realize the high potential of the site, we have initiated a project aimed at transforming the Zadko Observatory and surrounds into a Space Surveillance Hub in Western Australia that can be used by major international space agencies and commercial partners. 
The project is driven by a partnership with the European Space Agency (ESA), the Polish Space Agency (POLSA) and the Japan Aerospace Exploration Agency (JAXA). Each of these entities require observations of the Near-Earth environment at the specific location of the Zadko Observatory which are not currently feasible owing to the ageing of the observatory and its outdated components. Indeed, most of the electronic control systems within the Zadko Observatory were designed and manufactured in the late 1990’s and require urgent upgrades to ensure the longevity of this critically important facility. The existing computers cannot provide the level of cyber-security required by a modern observatory, and its instrumentation is becoming obsolete and degraded due to more than ten years of continuous operation. 

The global commercial investment in space-related technology is rapidly increasing \citep{gla00}. The Australian Government has recognized the economic benefits of this evolution, and is actively promoting links between defence, the space industry and the university sector. This investment outcome depends critically on the successful operation of space assets owned and operated by commercial entities and other international partners. One of the core topics covered by our current project is the possibility to track space hazards, allowing for pre-emptive actions such as an anticipated alert to a population, or planning an action to avoid collision and potential destruction of space and ground assets. Another topic is the possible use of the facility into the study of the degradation of satellites in orbit, in particular how payloads, structural components and solar panels react to space conditions over time, with vital information obtained used to assist space and satellite manufacturing commercial industry growth. In fact, this Project would enable a versatile Hub offering numerous benefits to the space sector.

Due to the aforementioned reasons, we have decided to upgrade the IT structure, critical to sustaining the activities of the Observatory, and the Telescope Control System (TCS) of the Zadko Telescope itself (including its motor, allowing for fast, precise tracking of satellites and Near-Earth Orbiters). These activities also involve an upgrade of the instrument chain, in order to increase the field of view of the telescope, in case of observations of objects with a poorly constrained position.

We do expect that these modifications will increase the request of observation time on the Zadko Telescope, due to its improved versatility, enabling further diversification of its scientific program. This will obviously also potentially lead to an over subscription of the telescope viewing time. For this reason, all requests for use of the Zadko Telescope will be handled by a scientific committee, granting access to the instrument based on the scientific merit of the request for 700 hours per year. This committee will also define a core program of 900 hours per year, based on collaborations signed between the Observatory and other research institutions, and a contributed program of 1000 hours per year, which will request a fee for observation time. This will provide additional funding for specific future improvements and upgrades. All of these programs will be open to any organization upon request. 

In addition, the Zadko Observatory has been able to attract commercial contracts and associated funding since 2016 due to its unique location and its expert support team. As part of the refurbishment process, climate control of the observation room and roof operations have been upgraded in order to maximize the roof open time, where any instrument can perform observations, while still allowing for some protection against inclement weather (see Gendre et al., these proceedings). These commercial contracts will continue with the potential for further development going forward. Some discussions have commenced with another entity for hosting of a new instrument. All the funds obtained by the commercial contracts and the paid contributed program using the Zadko Telescope will be used to support the ongoing day‐to‐day operation of the Zadko Observatory.

\section{Future plans}
\label{sec:future}

Having now secured consistent funding for the Zadko Observatory, considered critical for its ongoing operations and maintenance, we can now look to the future with confidence. As the nature of space situational awareness and space defence is rapidly changing, we consider ourselves to be in a transition phase to participate in this sector. One area that we have recently been invited to collaborate in, is the International Asteroid Warning Network \citep[IAWN,][]{onu14} with the aim to establish a worldwide network of instruments used to detect, track and physically characterize near-Earth objects (NEOs) to determine those that are potential impact threats to Earth. Consequently, in partnership with the University of New South Wales (UNSW), the JPL-NASA/CSIRO group based at Tidbinbilla and the University of Tasmania we have formed 'The Australian Consortium for Planetary Defence' using both optical and radar. The consortium of JPL-NASA/CSIRO/UNSW/UTAS/UWA, offers a wide baseline radio frequency and optical capability stretching from the West to East coast of Australia (Kruzins et al., in preparation). Another project we are about to commence in partnership with the Polish space agency (POLSA) is to understand how defunct satellites degrade over time to produce space debris. To accomplish this the project will employ low resolution spectroscopy to measure the surface degradation of geostationary satellites. 

As mentioned previously once the upgrades to the Zadko Telescope are complete we expect an increase in demand from various users. This increased use of telescope viewing time has several benefits being increased revenue, potential for collaboration, improved international visibility for the observatory plus maximum output from our scientific asset. Now referring to the site where the Zadko Observatory is situated as a 'Space Surveillance Hub' we are looking to expand beyond the Zadko Observatory to its surrounds. We are hosting another instrument for the USAFA's Falcon Telescope Network \citep{chu18} in close proximity to the Zadko Observatory and are currently in discussion with another international partner to host one other instrument onsite.    

\section*{ACKNOWLEDGEMENTS}
This research was supported by the Australian Research Council Centre of Excellence for Gravitational Wave Discovery (OzGrav), through project number CE170100004. E.M. acknowledges support from the Zadko Postgraduate Fellowship and International Space Centre (ISC).The Zadko Observatory is also supported by the University of Western Australia’s Office of Research and forms part of UWA’s International Space Centre (ISC).


\begin{thebibliography}
\bibitem[Chun et al.(2018)]{chu18} Chun, F.~K., Tippets, R.~D., Strong, D.~M., et al.\ 2018, \pasp, 130, 095003. doi:10.1088/1538-3873/aad03f
\bibitem[Coward et al.(2010)]{cow10} Coward, D.~M., Todd, M., Vaalsta, T.~P., et al.\ 2010, \pasa, 27, 331. doi:10.1071/AS09078
\bibitem[Coward et al.(2011)]{cow11} Coward, D.~M., Gendre, B., Sutton, P.~J., and Howell, E.~J. 2011, \mnras, 415, L26
\bibitem[Coward et al.(2017)]{cow17} Coward, D.~M., Gendre, B., Tanga, P., Turpin, D., Zadko, J., Dodson, R., Devogéle, M., Howell, E.~J. , Kennewell, J.~K., Boer, M., Klotz, A., Dornic, D., Moore, J.~A., and Heary, A. 2017, \pasa, 34, 16
\bibitem[Gladman et al.(2000)]{gla00} Gladman, B., Michel, P., and Froeschlé, C. 2000, Icarus, 146, 176
\bibitem[Moore et al.(2021)]{moo21} Moore, J.~A., Gendre, B., Coward, D.~M., et al.\ 2021, Revista Mexicana de Astronomia y Astrofisica Conference Series, 53, 35. doi:10.22201/ia.14052059p.2021.53.08
\bibitem[United Nations(2014)]{onu14} United Nations Office for outer space affairs, 2014, description available at https://iawn.net/documents/18-03593\_NEO\_Brochure\_Ebook.pdf

\end{thebibliography}
\end{document}